# CHAOS IN COMPUTING THE ENVIRONMENTAL INTERFACE TEMPERATURE: NONLINEAR DYNAMIC AND COMPLEXITY ANALYSIS OF SOLUTIONS

GORDAN MIMIĆ

*Faculty of Sciences, Department of Physics, University of Novi Sad,*
*Dositeja Obradovica Sq. 3, 21000 Novi Sad, Serbia*
*mimic.gordan@gmail.com*

DRAGUTIN T. MIHAILOVIĆ

*Faculty of Agriculture, University of Novi Sad,*
*Dositeja Obradovica Sq. 8, 21000 Novi Sad, Serbia*
*guto@polj.uns.ac.rs*

MIRKO BUDINČEVIĆ

*Faculty of Sciences, Department of Mathematics, University of Novi Sad,*
*Dositeja Obradovica Sq. 3, 21000 Novi Sad, Serbia*
*mirko.budincevic@dmi.uns.ac.rs*

ABSTRACT

We consider an environmental interface regarding as a complex system, in which difference equations for calculating the environmental interface temperature and deeper soil layer temperature are represented by the coupled maps. First equation has its background in the energy balance equation while the second one in the prognostic equation for deeper soil layer temperature, commonly used in land surface parameterization schemes. Nonlinear dynamical consideration of this coupled system includes: (i) examination of period one fixed point and (ii) bifurcation analysis. Focusing part of analysis is calculation of the Lyapunov exponent for a specific range of values of system parameters and discussion about domain of stability for this coupled system. Finally, we calculate Kolmogorov complexity of time series generated from the coupled system.

*Keyword*s: Chaotic behavior, energy exchange, coupled system, Lyapunov exponent, Kolmogorov complexity, domain of stability.

1. **INTRODUCTION**

Nowadays environmental sciences have a multidisciplinary approach to wide range of topics related to processes in biosphere, combining a lot of natural and even social disciplines. One of them is modeling processes through the environmental interfaces, which is a place where different parts of one system or different systems meet each other and exchange some kind of information.[1] Environmental interface is defined as *interface between two biotic or abiotic environments that are in relative motion and exchange energy, matter and information through physical, biological and chemical processes, fluctuating temporally and spatially regardless of space and time scale*.[2] There are a lot of examples of environmental interfaces in the nature such as interface between cells, human or animal bodies and surrounding environment, aquatic species and water around them and natural or artificially created surfaces and atmosphere.[3] A typical example of environmental interface in the nature is the ground surface, where exists all three mechanisms of energy transfer; incoming and outgoing radiation, convection of heat and moisture into the atmosphere and conduction of heat into deeper soil layers of ground.[4] Parameterization of these processes is of great importance for environmental models of different spatial and temporal scales.[5-7]

In Ref. 8 is shown that ground surface is treated as a complex system in which chaotic fluctuations occur while we compute its temperature. This system, as an actual dynamic system, is very sensitive to initial conditions and arbitrarily small perturbation of the current trajectory that may lead to its unpredictable behavior. In the aforementioned paper the lower boundary condition, i.e. the deeper soil layer temperature was constant, but it also can vary in time making with the energy balance equation a coupled system of equations. That system, often used in environmental models, is of interest to be analyzed by the methods of nonlinear dynamics.[9] Having in mind those facts, we: (i) perform a nonlinear dynamical analysis of coupled system for computing the environmental interface temperature and the deeper soil layer temperature and (ii) examine behavior of the coupled system in dependence on the main system parameters. In Section 2 we consider difference form of the energy balance equation and deeper soil layer temperature equation transforming them into the coupled system with the corresponding parameters. In Section 3 we analyze behavior of the solutions of the coupled system and we have examined domains of stability using Lyapunov exponent and Kolmogorov complexity. Conclusions are summarized in Section 4.

## 2. PHYSICAL BACKGROUND AND DERIVATION OF THE COUPLED SYSTEM

One of the most important conditions for functioning of any complex system is a proper supply of the system with energy. Dynamics of energy flow is based on the energy balance equation.[4] As we mentioned before, environmental interface is a complex system. General difference form of energy balance equation for the ground surface as an environmental interface is

$$C_g \frac{\Delta T_g}{\Delta t} = R_{net} - H - \lambda E - G \tag{1}$$

Where $T_g$ is the ground surface temperature, $\Delta t$ is the time step, $C_g$ is the soil heat capacity, $R_{net}$ is the net radiation, $H$ is the sensible heat flux, $\lambda E$ is the latent heat flux and $G$ is the heat flux into the ground. First, we assume that the net radiation is given as in Ref 10, i.e.

$$R_{net} = C_R (T_g - T_a) \tag{2}$$

where $T_a$ is the air temperature at some reference level and $C_R$ is the coefficient for the net radiation term. Second, we make expansion of the exponential term in the expression for latent heat flux

$$\lambda E = C_L d \left[ b(T_g - T_a) + \frac{b^2}{2}(T_g - T_a)^2 \right], \tag{3}$$

where $C_L$ is the water vapour transfer coefficient, $b = 0.06337\,°C^{-1}$, $d$ is parameter which occurs in expanding the series.[11] Further, the conduction of the heat into the soil can be written in the form

$$G = C_D (T_g - T_d), \tag{4}$$

where $C_D$ is the coefficient of the heat conduction while $T_d$ is the temperature of the deeper soil layer. The sensible heat flux $H$ can be parameterized as

$$H = C_H (T_g - T_a), \tag{5}$$

where $C_H$ is the sensible heat transfer coefficient. The prognostic equation for temperature of the deeper soil layer $T_d$ is

$$\frac{\Delta T_d}{\Delta t} = \frac{1}{\tau}(T_g - T_d), \tag{6}$$

where $\tau = 86400\,s$. After collecting all terms (2)-(6), the coupled system takes the form

$$C_g \frac{\Delta T_g}{\Delta t} = C_R (T_g - T_a) - C_H (T_g - T_a) - C_L d[b(T_g - T_a) \\ + \frac{b^2}{2}(T_g - T_a)^2] - C_D (T_g - T_d) \tag{7}$$

$$\frac{\Delta T_d}{\Delta t} = \frac{1}{\tau}(T_g - T_d). \tag{8}$$

More details about the nature and the range of physical parameters $C_R$, $C_L$, $C_D$ and $C_H$ can be found in Ref. 12. Now, using the time scheme forward in time (n indicates the time step) and dividing both sides of Eqs. (7) and (8) with the constant temperature $T_0$ (for example, value of mean Earth temperature, i.e. $T_0 = 288K$) we get

$$\frac{T_g^{n+1} - T_a^n}{T_0} = \frac{T_g^n - T_a^n}{T_0} + \frac{\Delta t}{C_g} C_R \frac{T_g^n - T_a^n}{T_0} - \frac{\Delta t}{C_g} C_H \frac{T_g^n - T_a^n}{T_0} - \frac{\Delta t}{C_g} C_L bd \frac{T_g^n - T_a^n}{T_0}$$
$$- \frac{\Delta t}{C_g} C_L dT_0 \frac{b^2}{2} \frac{(T_g^n - T_a^n)^2}{T_0^2} - \frac{\Delta t}{C_g} C_D \frac{T_g^n - T_a^n}{T_0} + \frac{\Delta t}{C_g} C_D \frac{T_d^n - T_a^n}{T_0} \quad (9)$$

$$\frac{T_d^{n+1} - T_a^n}{T_0} = \frac{T_d^n - T_a^n}{T_0} + \frac{\Delta t}{\tau} \frac{T_g^n - T_a^n}{T_0} - \frac{\Delta t}{\tau} \frac{T_d^n - T_a^n}{T_0} \quad . \quad (10)$$

Finally, introducing replacements $z = (T_g - T_a)/T_0$ and $y = (T_d - T_a)/T_0$, where $z$ is the dimensionless environmental interface temperature and $y$ is the dimensionless deeper soil layer temperature, we reach the coupled system

$$z_{n+1} = Az_n - Bz_n^2 + Cy_n \quad (11)$$
$$y_{n+1} = Dz_n + (1-D)y_n, \quad (12)$$

where $A = 1 + \frac{\Delta t}{C_g}(C_R - C_H - C_L bd - C_D)$, $B = C_L dT_0 \frac{b^2 \Delta t}{2C_g}$, $C = \Delta t \frac{C_D}{C_g}$ and $D = \frac{\Delta t}{\tau}$.

Introducing the replacement $z_n = Ax_n/B$, where $x$ is modified dimensionless environmental interface temperature and following Ref. 11 we can write

$$x_{n+1} = Ax_n(1 - x_n) + \frac{CB}{A} y_n \quad (13)$$

$$y_{n+1} = \frac{DA}{B} x_n + (1-D)y_n. \quad (14)$$

Analysis of values of parameters $A$, $B$, $C$ and $D$, based on a large number of energy flux outputs from the land surface scheme runs, indicates that their values are ranged in the following intervals: (i) $A \in [0, 4]$ and (ii) $B$, $C$ and $D$ are ranged in the interval [0,1]. Thus, A is the logistic parameter, which from now will be denoted with $r$. All other groups of parameters in the system (13)-(14) have the values in the same interval [0,1]. Let us underline that under some circumstances those parameters can be equal. Correspondingly, we replaced all of them by introducing the coupling parameter $\varepsilon$.

Finally, system (13)-(14) can be written in the form of coupled maps, i.e.,

$$x_{n+1} = rx_n(1 - x_n) + \varepsilon y_n \qquad (15)$$

$$y_{n+1} = \varepsilon(x_n + y_n) \qquad . \qquad (16)$$

## 3. NONLINEAR DYNAMICAL ANALYSIS

We now examine the effect of coupling two nonlinear maps given by Eqs. (15)-(16), where are the logistic parameter $r \in [0,4]$ and the coupling parameter $\varepsilon \in [0,1]$. This map displays a wide range of behaviour as the parameters $r$ and $\varepsilon$ change. We consider system of difference equations of the form $X_{n+1} = F(X_n)$ with notation $F(X_n) = (rx_n(1-x_n) + \varepsilon y_n, \varepsilon(x_n + y_n))$, where $X_n = (x_n, y_n)$ is a vector representing the dimensionless environmental interface temperature and the deeper soil layer temperature, respectively. We look for the fixed point of mapping given by (15)-(16) using criterion $X = F(X)$. Thus, we get $(0,0)$ and $\left((r + \varepsilon^2/(1-\varepsilon) - 1)/r, \varepsilon/(1-\varepsilon)[(r + \varepsilon^2/(1-\varepsilon) - 1)/r]\right)$ as two fixed points. Now, for the fixed point $(0,0)$ we have two eigenvalues $\lambda_{1,2} = (r + \varepsilon \pm \sqrt{r^2 - 2r\varepsilon + 5\varepsilon^2})/2$. Using the one with the plus sign, which has higher absolute value, and the criterion that fixed point is attractive if $|\lambda| < 1$ and it is repulsive if $|\lambda| > 1$, we localize regions in $(\varepsilon, r)$ plane which tell us for what pair of parameter values fixed point $(0,0)$ would be either attractive or repulsive. Applying the same procedure for the other fixed point given by

$$\left((r+\varepsilon^2/(1-\varepsilon)-1)/r, \varepsilon/(1-\varepsilon)[(r+\varepsilon^2/(1-\varepsilon)-1)/r]\right)$$ and with the eigenvalues

$$\lambda_{3,4} = \frac{1}{2(\varepsilon-1)}(-2+\varepsilon+3\varepsilon^2+r-\varepsilon r) \pm \sqrt{(2-\varepsilon-3\varepsilon^2-r+\varepsilon r)^2 - 4(\varepsilon-1)(-2\varepsilon+3\varepsilon^2+\varepsilon^3+\varepsilon r-\varepsilon^2 r)}$$

we get exactly the same regions of attraction and repulsion in (ε,r) plane as it is depicted in Fig. 1.

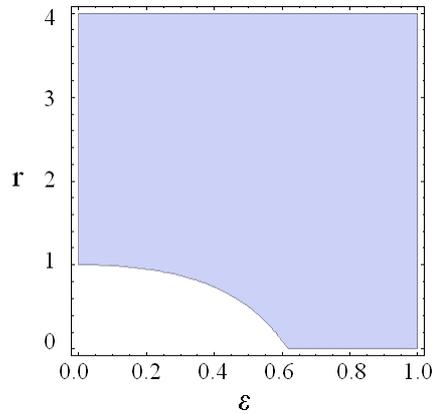

Fig. 1. Graphical interpretation of fixed points for the coupled maps (15)-(16) as a function of logistic parameter $r$ and coupling parameter $\varepsilon$. Both fixed points are in the following regions: (i) attractive (white) and (ii) repulsive (grey).

Bifurcation diagrams for *x* and *y* are given in Fig. 2 as a function of the logistic parameter *r* and for this coupled maps were plotted with *r* ranging from 0 to 4 and for $\varepsilon = 0.1$. For each value of *r*, we used the final point of the previous *r* value and then 500 iterations are plotted. We noticed that maximum values of *y* strongly depend on coupling parameter $\varepsilon$. Let us note that value of coupling parameter is small. Thus, bifurcation diagram of *x* is close the logistic map. Bifurcations start after the parameter *r* reaches value 3 and chaotic regime exists after *r*=3.5 on the both diagrams.

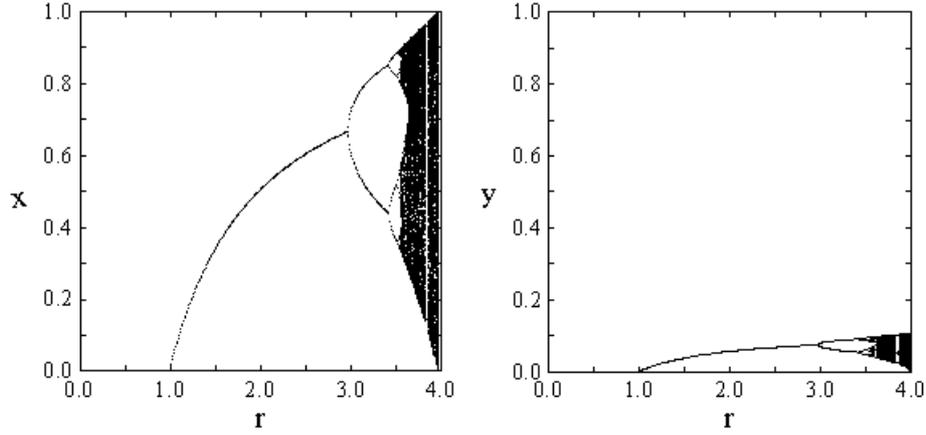

Fig. 2. Bifurcation diagrams for the coupled maps (15)-(16) with $r$ ranging from 0 to 4 and $\varepsilon = 0.1$. Initial conditions were $x_0 = 0.2$ and $y_0 = 0.4$.

We have also plotted the phase diagram for $x$ and $y$, which is depicted in Fig. 3. This plot was obtained by iterating $x$ (from 0 to 1) and $y$ (from 0 to 0.15). In those calculations, 1000 iterations were applied for the initial state ($x_0 = 0.2$, $y_0 = 0.4$) after 200 steps of stabilization of $(x, y)$ pair. From this figure it is seen that this plot is similar to Henon's attractor.[13] It was expected, because for $D = 1$ dynamical system given with Eqs. (15)-(16) is similar to Henon map.

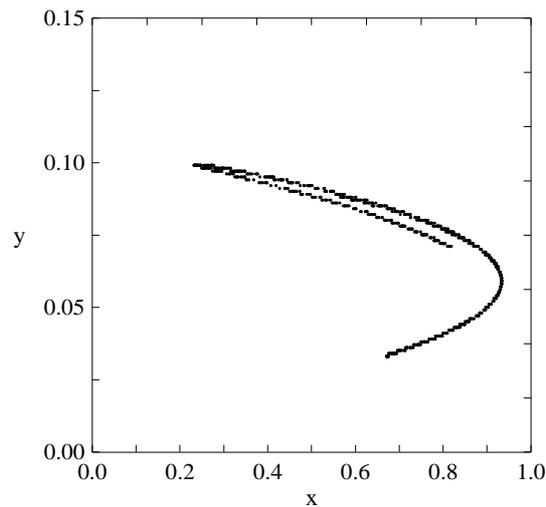

Fig. 3. Phase diagram of the map (15)-(16) for $r = 3.7$, $\varepsilon = 0.1$, and initial point $x_0 = 0.2$, $y_0 = 0.4$.

Irregularities in solution of the system (15)-(16) can come from two reasons. They are: (i) numerical, i.e. because we try to choose appropriate difference equation whose solution is "good" approximation to the solution of the given differential equation[14] and (ii) physical, i.e. occurrence of chaotic fluctuations in the considered system because the environmental interface cannot oppose an enormous radiative forcing, suddenly reaching the interface (Fig. 4). Therefore, the crucial point is whether the time scale of discretization is appropriate for the considered physical phenomenon. However, the energy balance equation we analyzed in difference form captures time scale of the energy exchange on the environmental interface ranged from very small to the large scale in time, but also in space. In the simulation example with LAPS scheme in Fig.4, the time step that we have used was 1800 s for which we got range of parameters in coupled system (15)-(16). Note that this time step is quite realistic for use in many environmental models.

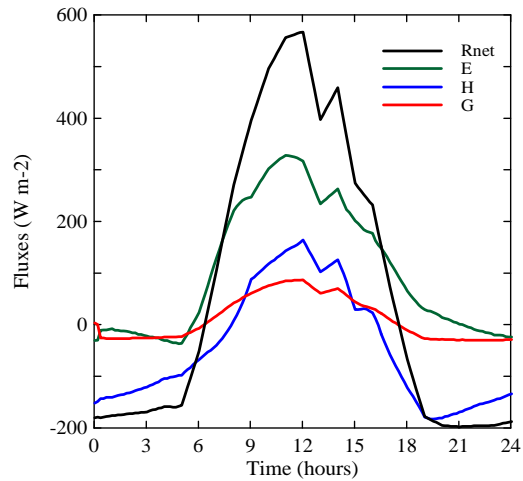

Fig. 4. Diurnal cycle of energy balance equation components. Second peak in net radiation term can be noticed in all other fluxes. Simulation was done using LAPS.[15]

Let us note that the assumption $T_g, T_d \geq T_a$ is violated in dependence on atmospheric conditions. However, there exist conditions for which this criterion is satisfied since the ground surface ("skin") temperature can be even for 10 °C higher then air temperature at 2 m hight.[4] We wanted to point out that in these kind of situations there is a possibility for occurrence of chaotic phenomena, which can cause uncertainties in calculations of the ground surface temperature. This is because of drawback of currently designed environmental models to calculate the ground surface temperature under these conditions. Looking at the system (15)-(16) we have to keep in mind some a priori mathematical limitations. As we take the range of x, y between 0 and 1, it is seen from Eq. (16) that parameter $\varepsilon$ has to be less or equal to 0.5. Further, from the Eq. (15), where maximum value for *x* is 0.5 and for *y* is 1, we get new condition that $r/4+\varepsilon \leq 1$. Because of Eq. (15) has form of logistic map we know that chaos is present in a case when logistic parameter *r* is in interval [3,4] so parameter of coupling $\varepsilon$ should be very small, i.e. beyond 0.1. Thus, selection of initial conditions is also very important in evolution of system.

Therefore, it raises the question whether we can find either domain or domains where physically meaningful solutions exist. We do that by considering the stability of physical solutions, using Lyapunov exponent, which is a measure of convergence or divergence of near trajectories in phase space. Sign of Lyapunov exponent is characteristic of attractor type and for stable fixed point is negative, although for chaotic attractor is positive. We analyze behaviour of Lyapunov exponent for the corresponding autonomous equation which uniformly changes in intervals of *r* and $\varepsilon$. For our system $X_{n+1}=F(X_n)$, $X_n=\begin{pmatrix}x_n\\y_n\end{pmatrix}$ and for any initial point $(x_0,y_0)$ from attracting region, characterization of asymptotic behavior of the orbit[16-17] is given by the largest Lyapunov exponent,[18] presented by

$$\lambda = \lim_{n\to\infty}\left(\ln\left\|\prod_{s=1}^{n}\xi_s\right\|/n\right) \tag{17}$$

where $\xi_s$ is Jacobian matrix, given by

$$\xi_s = \begin{bmatrix} r(1-2x_s) & \varepsilon \\ \varepsilon & \varepsilon \end{bmatrix}. \tag{18}$$

Calculating Lyapunov exponent for the coupled system (15)-(16) with values of parameters $\varepsilon \in (0.05,0.1)$ and $r \in (3.6,3.8)$ we got results depicted in Fig. 5. It is shown that Lyapunov exponent mostly has positive values which approve presence of chaos in this system, but there are still some strait regions where the Lyapunov exponent is negative and where the solutions of the coupled system are stable, i.e. domains of stability.

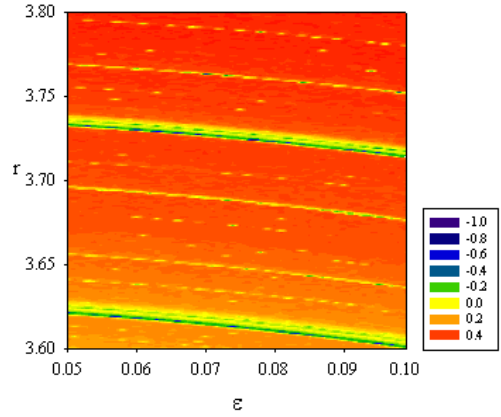

Fig. 5. Lyapunov exponent of the coupled system (15)-(16), which shows presence of strait regions of stability in highly developed chaos.

For nonlinear time series analysis we use Kolmogorov complexity, interpreted by Lempel and Ziv.[19] This measure of complexity estimates degree of disorder or irregularity in sequence, counting the repeating subsequences of all lengths. First we encode time series by forming sequence S made of characters 0 and 1 using the rule

$$S(i) = \begin{cases} 0, & x_i < x_* \\ 1, & x_i \geq x_* \end{cases} \quad (19)$$

where $x_*$ is threshold, and in our case threshold is mean value of the time series. Now we calculate the complexity counter $c(n)$ which is defined as minimum number of distinct patterns contained in sequence S,[20] and its ultimate value $b(n)$ is given by

$$b(n) = \frac{n}{\log_2 n} \quad (20)$$

Finally, normalized complexity measure $C_k(n)$ is defined as

$$C_k(n) = \frac{c(n)}{b(n)} = c(n)\frac{\log_2 n}{n} \quad (21)$$

and for nonlinear time series it has values between 0 and 1. Again, the idea was to calculate Kolmogorov complexity of time series produced for chaotic states of coupled system with range of parameters $\varepsilon \in (0.05, 0.1)$ and $r \in (3.6, 3.8)$.

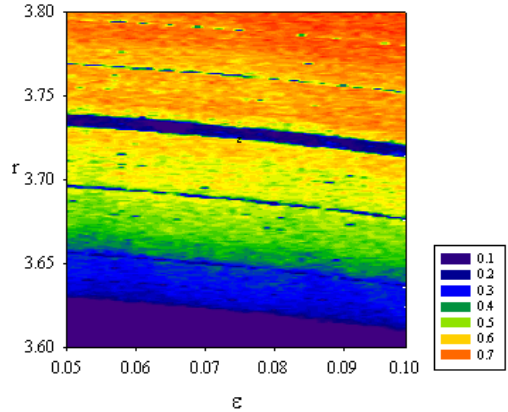

Fig. 6. Kolmogorov complexity for the coupled system of Eqs. (15)-(16) as a function of parameters *r* and $\varepsilon$.

It is seen from Fig. 6 that complexity of the system strongly depends on parameter *r*. Higher value of Kolmogorov complexity implies highly developed chaos. Although, there are still regions, coloured with purple, which are related to domains of stability with a stable solution and nonchaotic behavior of system.

## 4. CONCLUSIONS

We derived, from the energy balance equation, a coupled system of difference equations for calculating the dimensionless environmental interface temperature and deeper soil layer temperature to analyze environmental interface as complex system. We have performed the nonlinear dynamical analysis, which includes (i) determination of the period one fixed point and bifurcation diagrams, (ii) calculation of the Lyapunov exponent and (iii) we have discussed domains of stability of the solutions for this system in a specific interval of parameters values. The results showed that behaviour of system is highly dependent on values of parameters *r* and *ε*. Finally, we have calculated Kolmogorov complexity as a measure of nonlinearity of the system. We point out on the fact that there exist conditions when the environmental interface temperature can not be calculated by the physics of currently designed environmental models, because of occurrence of chaotic phenomena on the interface.

**Acknowledgements**

This paper was realized as a part of the project "Studying climate change and its influence on the environment: impacts, adaptation and mitigation" (III43007) financed by the Ministry of Education and Science of the Republic of Serbia within the framework of integrated and interdisciplinary research for the period 2011-2014. The authors are grateful to the Provincial Secretariat for Science and Technological Development of Vojvodina for the support under the project "Climate projections for the Vojvodina region up to 2030 using a regional climate model" funded by the contract No. 114-451-2151/2011-01.